\documentclass[12pt]{iopart}
\usepackage{graphicx} 
\usepackage{caption}
\usepackage{subcaption}
\usepackage{cite} 

\begin{document}

\title[Inverse GPE problem with ANN ]{Solution of inverse problem for Gross-Pitaevskii equation with artificial neural networks}
\author{S P Pokatov$^1$, T Yu Ivanova$^1$, D A Ivanov$^{2,1}$}
\address{$^1$Saint-Petersburg State University, Ulianovskaya 3, Saint-Petersburg, Russia}
\address{$^2$Soft-Impact Ltd, Bolshoy Sampsonievskiy 64, Saint-Petersburg, Russia}

\vspace{10pt}
\begin{indented}
    \item[]June 2023
\end{indented}

\begin{abstract}
    We propose an Artificial Neural Network (ANN) design to solve the inverse problem for a 1D Gross-Pitaevskii equation  (GPE). More precise, the ANN takes the squared modulus of the stationary GPE solution as an input and returns the parameters of the potential function and the factor in front of the GPE non-linear term. From the physical point of view the ANN predicts the parameters of a trap potential and the interaction constant of 1D Bose-Einstein Condensate (BEC) by its density distribution. Using the results of numerical solution of GPE for more than $30 000$ sets of GPE parameters as train and validation datasets we build the ANN as a fast and accurate inverse GPE solver. 
\end{abstract}


\section{Introduction}
Ultra cold gases are very promising objects for wide range of practical and fundamental applications. These applications range from quantum metrology to quantum information processing~\cite{metrology1, metrology2, info-processing}. The peculiar feature distinguishing degenerate quantum gases~\cite{wiseman, hope1, hope2, ivanov1, ivanov2, ivanov3} from thermal ensembles is the possibility to observe quantum superposition on the macroscopic level. Thus the methods of control degenerate quantum gases without disturbing their "quantumness" are of great interest. The control protocols especially those designed for the real time applications should be fast enough to minimize the environmental noise effects. It is natural that such protocols involve some sort of feedback control based on the measurement of the gas density distribution. The extraction of the information about the system is the key component of any feedback control scheme. Thus a practical method to effectively extract the information on the cold atomic ensemble is of great importance.    

In the realm of classical systems the promising tool to help to extract the information on the system dynamics is Artificial Intelligence and Machine Learning (ML). The application of ML to various problems of experimental and theoretical atomic and molecular physics is becoming more and more popular \cite{Montavon-molecular-electronic-properties, Yao-kinetic-energy, Caetano-diff-eq-in-atomic-phys, Li-predict-y-no-bond-homolysis, Monterola-nonlinear-schrodinger}. The ML methods proved their efficiency in optimizing the application of Density Functional Theory \cite{Snyder-density-func, Brockherde-kohn-sham-eq, Rupp-molecular-atomization-energies, Seko-density-func-theory, Behler-pot-energy-surfaces} and solution of Schr{\"o}dinger equation \cite{Mills-Schrodinger-eq}. 

The ML methods have been applied to describe ensembles of atoms in lattices and trapped Bose-Einstein Condensates. In \cite{Liang-Generation-BEC} the authors demonstrate the Artificial Neural Network (ANN) trained to solve Gross-Pitaevskii equation (GPE). It has been shown that the accuracy of such a solver is very high while the time required to get the solution is negligible compared to the time required to numerically solve GPE on a typical modern computer. Among multiple possible applications this ANN GPE solver can be used as a part of a real time control system.

However, it is even more important from the practical point of view to be able to solve the inverse GPE problem. That is to find the potential or the interaction constant having the knowledge of the BEC density distribution. The inverse GPE solver can help to extract the information from various sensors based on BEC and thereby increase their accuracy. There are multiple possible applications of such sensors in geophysics, metrology and tests of fundamental physics \cite{Geiger-measure-with-cold-atom-sensors, Peters-measure-gravit-acceleration, Debs-gravimetry_with_BEC}.  In particular, the setups based on BEC have been proposed for magnetic sensors \cite{Pelegri-sensing-BEC-modes} and gravitational wave detectors \cite{Sabin-phonon-creation-by-grav-waves}. Furthermore, the possibility of quick solution of the inverse problem can help to improve the time resolution of the sensors. 

One can claim that the solution of the mentioned inverse problem is only meaningful if the density distribution of BEC can be one-to-one mapped to the set of parameters of GPE. Unfortunately, the rigorous mathematical prove or disprove of this property of the mapping is unknown to the authors. The results presented below can be considered as a practical way to address this question. Also the conclusions we made below should not be extrapolated to arbitrary potentials, but should be restricted to the cases with certain $\textit{a priori}$ knowledge of the potential shape. In our research we restrict ourselves to symmetric double-well potentials. 

Bearing the disclaimer of the previous paragraph in mind the paper demonstrates the feasibility of ANN to solve the inverse GPE problem. In particular, it is shown that an ANN can be designed and trained to accurately reconstruct the parameters of a double-well potential and the interaction constant of BEC using the density distribution as an ANN input. 

\section{Model}
The common mathematical tool to describe the steady state of an one-dimensional atomic BEC in an external potential $V_{ext}(x)$ is GPE for the condensate wavefunction $\psi (x)$:
\begin{equation}
    \left(-\frac{\hbar^2}{2m} \frac{d^2}{dx^2} + V_{ext}(x) + g|\psi(x)|^2\right)\psi(x)=\mu\psi(x).
    \label{eq:GPE}
\end{equation}
In this equation the mass of an individual atom is $m$ while the Planck's constant is $\hbar$. The low-energy collisions of atoms are described via the interaction constant $g$. The right hand side of~(\ref{eq:GPE}) contains the chemical potential $\mu$ of the considered ensemble. Contrary to commonly used normalization we require
\begin{equation}
    \label{eq:bwc-norm}
    \int dx |\psi (x)|^2 = 1.
\end{equation}
Thus, the BEC wavefunction describes not the density, but the probability density of the trapped atoms. 

For our purpose we need  a non-trivial potential that can be characterized by only a few parameters. Otherwise the training dataset could be huge and the training process can become too time consuming. We chose the symmetric double-well external potential $V_{ext}(x)$ as it is frequently discussed in the literature with respect to different possible applications. The potential function is then given by 
\begin{equation}
V_{ext}(x)=8V_{min}\left(2\frac{x^4}{\xi^4}-\frac{x^2}{\xi^2}\right),
\label{eq:dw-potential}
\end{equation}
where $V_{\min}$ is the depth of the potential well, $\xi$ is the distance between the potential minima. These two parameters have transparent physical meaning and apart from a gauge constant completely characterize the potential functions. The ANN discussed below is designed and trained to predict $V_{min}$ and $\xi$ as well as the interaction constant $g$ using the BEC density values $|\psi(x)|^2$ in certain sample points as an input.

\section{Data Collection and ANN topology}
The key element of the success of ANN is the availability of high quality labeled data. For our purpose we generate data by solving GPE with various potential parameters and interaction constants. The GPE is numerically solved using the Julia-based free framework QuantumOptics.jl~\cite{QuantumOptics}. Since the solution time for a single case is quite small it is possible to generate density distributions for quite a dense grid of the problem parameters. Thus, there is no need in Design of Experiments methods~\cite{DOE}. 

The GPE was solved for the coupling strengths $g$ in the range of $1000\!-\!10800$ with the step of $200$. For each $g$ the depth of the potential well $V_{min}$ ranges in $100\!-\!2500$ with the step of $100$. For each pair of $g$ and $V_{min}$ the distance between the potential minima $\xi$ covers the range $5.0\!-\!9.8$ with the step of $0.2$. The units of the parameters are derived from the used convention $\hbar = 1$ and $m = 1$. The interaction constant $g$ here differs from the usually used definition by the factor of the total number of atoms. Thus the complete dataset contains $50 \times 25 \times 25 = 31250$ entries. The $70\%$ of this dataset is used to train the model and the rest is reserved for the evaluation of ANN prediction accuracy.

\begin{figure}
    \centering
    \includegraphics[width=\textwidth]{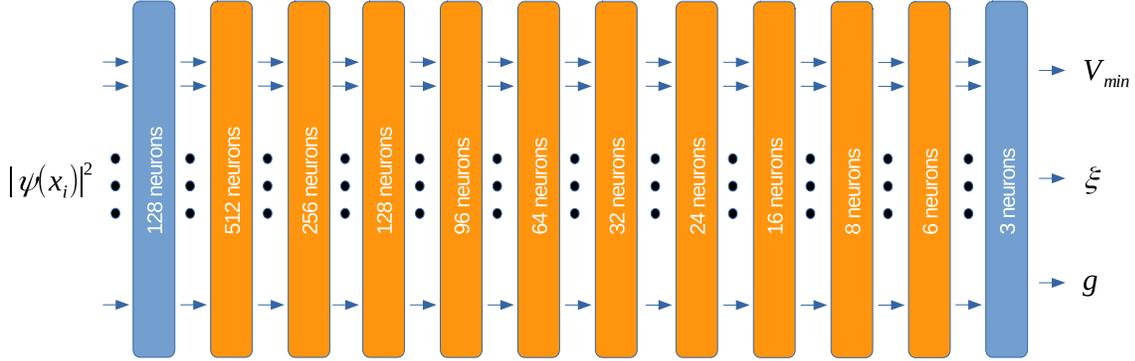}
    \caption{Schematic representation of the working ANN model. The ANN is fully connected with 10 hidden layers shown in orange. The input layers are shown in blue.}
    \label{fig:ann-model}
\end{figure}

The solution of GPE is performed on the grid of $128$ points using the imaginary time propagation technique~\cite{ITPT}. The input of the ANN accepts the density values on the uniform grid of points $x_i$, $i = 1..128$. Thus, the input layer of ANN contains $128$ neurons. Since ANN is designed to predict $3$ parameters there are $3$ neurons in the output layer as shown in figure~\ref{fig:ann-model}.   

Being limited to 1D case and moderate ($128$) number of input neurons we decided to use the fully connected network. Such ANN can result in more accurate predictions provided sufficient amount of data as compared with, for example, a convolution one.

We tested multiple designs of ANN and eventually came up with the model shown in figure~\ref{fig:ann-model}. The ANN contains $10$ hidden layer shown in orange with the following numbers of neurons: $512$, $256$, $128$, $96$, $64$, $32$, $24$, $16$, $8$, $6$. These values were manually adjusted until the work of ANN met predefined criteria. The activation function for all neurons is LeakyReLU.

The ANN model was implemented with Python using the Keras TensorFlow library~\cite{keras}. To train the model we applied Adam algorithm with the initial learning rate $0.001$. This value was found to provide reasonable time of convergence. Following the strategy of stochastic gradient decent methods the whole training dataset was divided into the number of mini-batches with $2000$ elements in each mini-batch.

\begin{figure}
    \centering
    \includegraphics[width=0.5\textwidth]{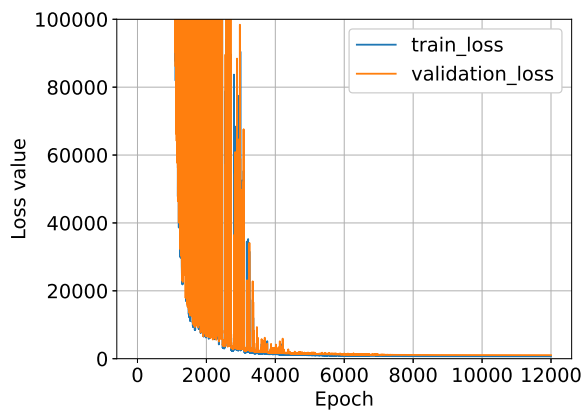}
    \caption{Train and validation errors versus the number or train epochs.}
    \label{fig:learning}
\end{figure}

The process of training the model is shown in figure~\ref{fig:learning}. It demonstrates rapid oscillations of the loss function on the validation set in the beginning of the learning process. The loss function is defined as
\begin{eqnarray}
\label{eq:loss}
\mathrm{LF}\left(g^{(pr)}, \xi^{(pr)}, V_{min}^{(pr)}\right) = \nonumber \\ \frac{1}{N_{mb}} \sum_{n} \left[\left(g_n - g^{(pr)}\right)^2 + \alpha\left(\xi_n - \xi^{(pr)}\right)^2 + \beta\left(V_{min, n} - V_{min}^{(pr)}\right)^2\right],
\end{eqnarray}
where $N_{mb}$ is the size of the mini-batch; $g^{(pr)}$, $\xi^{(pr)}$ and $V_{min}^{(pr)}$ are current predictions of the ANN. In order to increase the weight of the parameters with smaller absolute values the weighting coefficients $\alpha$ and $\beta$ are introduced.

As it is seen from figure \ref{fig:learning} the reasonable convergence requires more than approximately $6000$ training epochs. To speedup the learning the Early Stopping technique was used. This algorithm controls the value of the loss function and decreases the learning rate if the loss value stabilizes. If decreasing of the learning rate does not help the Early Stopping algorithm terminates training.

\section{Results}
\begin{figure}
	\centering
	\includegraphics[width=0.5\textwidth]{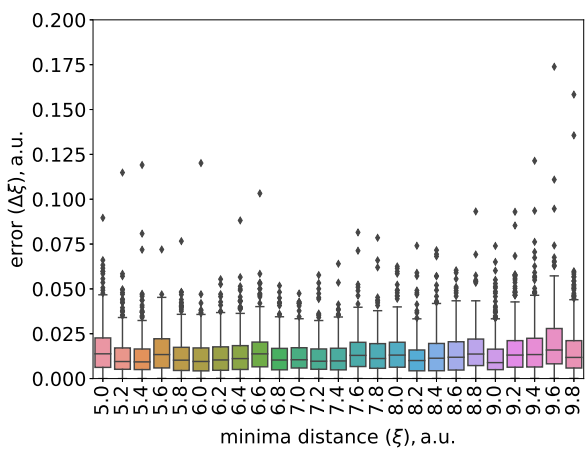}
	\caption{Box plot of prediction error for minima distance $\xi$.}
	\label{fig:error-min-sep}
\end{figure}

\begin{figure}
	\centering
	\includegraphics[width=0.5\textwidth]{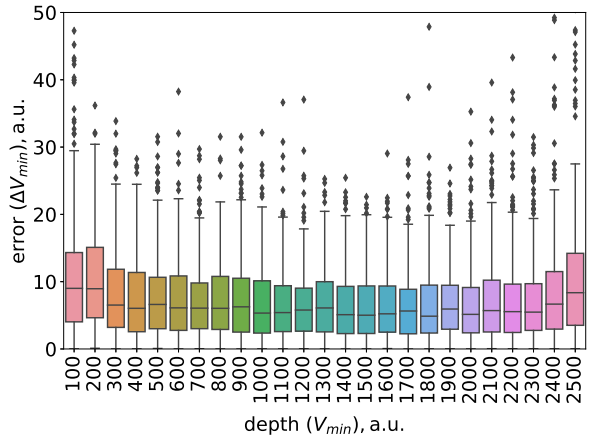}
	\caption{Box plot of prediction error for potential depth $V_{min}$.}
	\label{fig:error-depth}
\end{figure}

\begin{figure}
    \centering
    \includegraphics[width=0.75\textwidth]{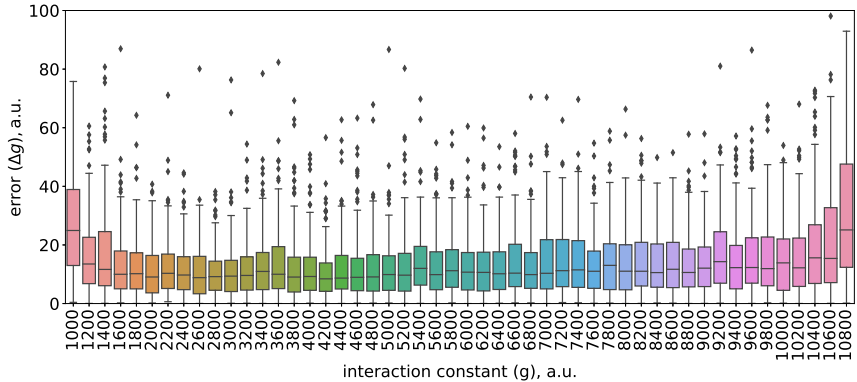}
    \caption{Box plot of prediction error for interaction constant $g$.}
    \label{fig:error-g}
\end{figure}

After training the ANN we tested its predictions on the dataset of $30\%$ of the whole available data. The box plots for the prediction errors are shown in figures~\ref{fig:error-min-sep},~\ref{fig:error-depth} and~\ref{fig:error-g}. Here, the absolute values of the errors and their statistics are addressed to indicate that the errors are weakly dependent on the absolute value of the predicted parameter. Thus, the larger parameters will be predicted with better relative accuracy. The horizontal lines inside boxes show the $0.5$ quantile (median), the lower (upper) edges of the boxes indicate $0.25$ ($0.75$) quantile. The length of the "whiskers" is $1.5$ of the box sizes (inter quartile range) if there are data points outside of this range otherwise the length indicates the position of the most distant point. 

Another visible trend that is especially pronounced in figures~\ref{fig:error-depth} and~\ref{fig:error-g} is the increase of the prediction error near the lower and the upper limits of the parameter variation range. This seems natural since the ANN does not get enough information on the effect of the parameter variation at the limits of the range. However, for practical application one should take this effect into account and use the ANN predictions with cautions if they are close to the limits of the training parameter ranges. 

Apart from this trend, in the most cases ($75\%$) the prediction error for the minima separation $\Delta \xi < 0.025$, for the potential depth it is $\Delta V_{min} < 15$, for the interaction constant $\Delta g < 20$. Thus, in the worst case corresponding to low values of $V_{min}$ the prediction error is estimated as about $15\%$. This clearly demonstrates the possibility of using ANN to predict the GPE parameters from the density distribution. 

\begin{figure}
	\centering
	\includegraphics[width=0.5\textwidth]{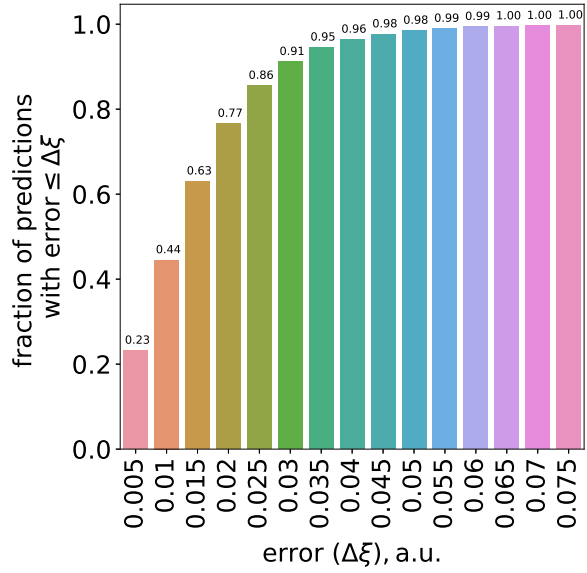}
	\caption{Statistics of the minima distance error $\Delta \xi$.}
	\label{fig:min-sep-color}
\end{figure}

\begin{figure}
	\centering
	\includegraphics[width=0.5\textwidth]{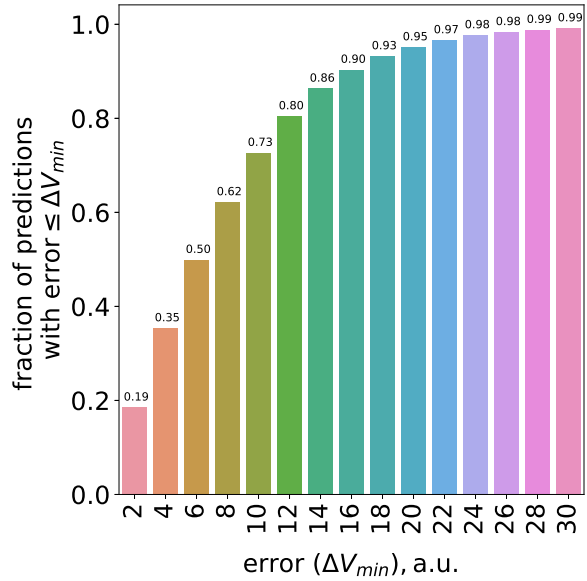}
	\caption{Statistics of the potential depth error $\Delta V_{min}$.}
	\label{fig:depth-color}
\end{figure}

\begin{figure}
	\centering
	\includegraphics[width=0.5\textwidth]{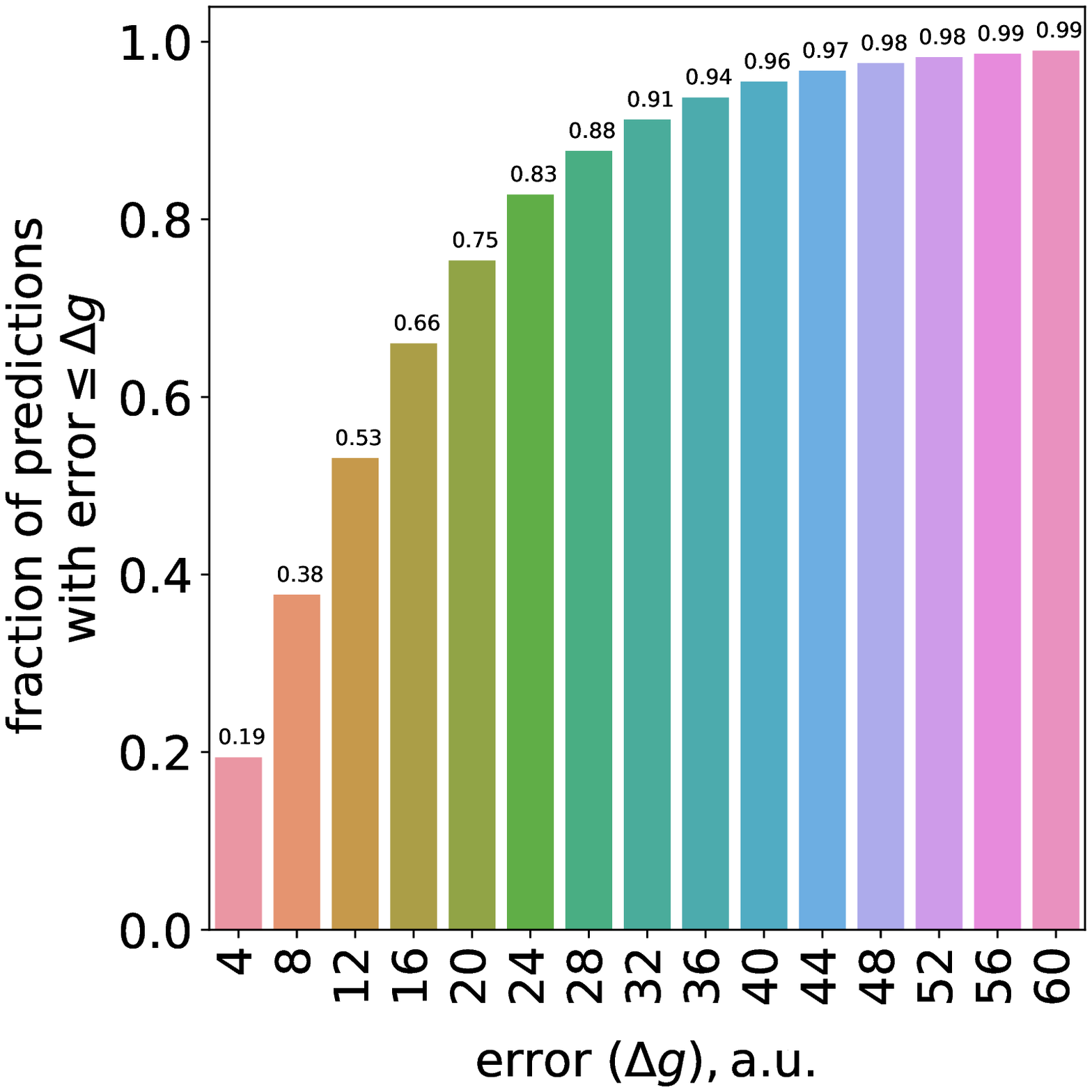}
	\caption{Statistics of the interaction constant error $\Delta g$.}
	\label{fig:g-color}
\end{figure}

Another informative way to describe the efficiency of the designed ANN is using the statistics of the errors, that ignores the values of the predicted parameters. The bar charts representing the fraction of the predictions with error smaller than the given value are shown in figures ~\ref{fig:min-sep-color},~\ref{fig:depth-color} and~\ref{fig:g-color}. The $95\%$-confidence interval for the minima separation is $0.035$, which is less than $1\%$ of the smallest value in the dataset. The confidence interval for the potential minimum is $16$. This is $16\%$ of the minimal value in the dataset. For the interaction constant the confidence interval is $32$, which is about $3.2\%$ of the minimal value in the dataset.  

These plots can be interpreted as a probability of getting  the error below the certain value regardless what value is predicted. In all the cases this probability rapidly grows for small errors and grows slowly for larger errors. The error value that distinguishes these different trends approximately corresponds to the green area in the figures~\ref{fig:min-sep-color},~\ref{fig:depth-color} and~\ref{fig:g-color}.

Note that the prediction of different parameters demonstrates slightly different behavior as the error value grows. In particular, the prediction of the minima separation $\xi$ shows faster growth for small errors and slower growth of the probability for larger errors than the prediction of the potential depth $V_{min}$ and the interaction constant $g$. Thus, the prediction of the potential minima separation is easier for the ANN that the prediction of other parameters. Interestingly, the same seems to be true for a human who can quantitatively predict the separation by simply measuring the distance between the BEC density peaks. However, the depth of the potential and the interaction constant can only be qualitatively estimated by looking at the width of the BEC density distribution.  

\section{Conclusions}
We designed and trained the ANN that predicts the potential parameters and the interaction constant of a trapped BEC based on its density distribution. The training and validation datasets were generated by the numerical solution of stationary GPE on a 1D grid of spacial points. 

The fully connected ANN with $10$ hidden layers was trained on a set of about $30000$ data samples. The largest prediction error is obtained for small values of the potential depth. It is $16\%$. The accuracy of the prediction of other parameters in all cases is better than $3.2\%$.  This clearly demonstrates the feasibility of the used approach and indicates the possibility of using it in practical applications.

It was found that the prediction of the distance between the minima of a double-well potential is in some sense easier for the ANN than the predictions of other parameters. Intuitively this can be understood by the fact that the distance between the minima has a more straightforward quantitative relation to an easily extracted distance between the peaks of the BEC density distribution. 

\section{Acknowledgment}
We thank Vera Baturo from the department of Photonics in Saint-Petersburg University for useful discussions and comments.

\end{document}